\documentclass[a4paper]{article}
\usepackage{Odyssey2024}
\usepackage{epsfig,amssymb,amsmath}
\usepackage{multicol}
\usepackage{subcaption}
\usepackage{multirow}
\usepackage{float}
\usepackage{booktabs}
\usepackage{cite}
\usepackage{hyperref}
\usepackage{xcolor}
\ninept

\title{Exploring speech style spaces with language models: \\
Emotional TTS without emotion labels}

\name{Shreeram~Suresh~Chandra, Zongyang Du, Berrak Sisman}

\address{Speech \& Machine Learning Lab, The University of Texas at Dallas, TX, USA }
\begin{document}
\maketitle

\begin{abstract}

Many frameworks for emotional text-to-speech (E-TTS) rely on human-annotated emotion labels that are often inaccurate and difficult to obtain. Learning emotional prosody implicitly presents a tough challenge due to the subjective nature of emotions. In this study, we propose a novel approach that leverages text awareness to acquire emotional styles without the need for explicit emotion labels or text prompts. We present TEMOTTS, a two-stage framework for E-TTS that is trained without emotion labels and is capable of inference without auxiliary inputs. Our proposed method performs knowledge transfer between the linguistic space learned by BERT and the emotional style space constructed by global style tokens. Our experimental results demonstrate the effectiveness of our proposed framework, showcasing improvements in emotional accuracy and naturalness. This is one of the first studies to leverage the emotional correlation between spoken content and expressive delivery for emotional TTS.\footnote{Demo page:  https://kodhandarama.github.io/TEMOTTSdemo}
\end{abstract}


%
\section{Introduction}
Recent neural text-to-speech (TTS) models \cite{Wang2017TacotronTE,ren2021fastspeech,lancucki2021fastpitch} have demonstrated the capability to generate high-quality speech when provided with a large amount of high-quality data, allowing for satisfactory alignments between speech and text. As we aim to bridge the gap between human speech and synthesized speech, we recognize that prosody modeling has a huge role to play in the synthesis of natural and expressive speech \cite{liu2021expressive}. Prosody modeling in deep neural TTS models has been achieved by predicting low-level features such as duration, pitch, energy, and emphasis. However, the synthesis of emotional speech remains a challenging task due to the highly complex nature of emotions. 

The goal of E-TTS systems is to effectively capture the prosodic aspects of speech, enabling the accurate conveyance of emotion-specific paralinguistic information to the listener. Emotions are highly stochastic functions of low-level prosody features - which makes it hard to explicitly model emotional prosody. This task is also constrained by the limited availability of emotion-annotated datasets. E-TTS systems have the added benefit of being emotion-aware compared to expressive TTS models, thereby providing the opportunity for new applications like personalized virtual assistants, interactive storytelling systems, and emotive AI therapists.

Previous approaches in E-TTS have predominantly relied on emotion labels to condition models for learning emotional speaking styles. Early approaches used one-hot encoded emotion embeddings during training \cite{lee2017emotional}, while later work popularized the use of global style tokens (GST) \cite{lee2017emotional}. Various GST-based methods such as centroid-based weights \cite{Kwon2019AnES}, interpolation algorithms \cite{Um2019EmotionalSS}, and joint style-linguistic emotion embeddings \cite{Kwon2019EmotionalSS} were applied to generate emotion-specific styles. These methods, however, encourage the style encoder to learn fixed style embeddings, thereby limiting exploration of the style space. In \cite{Liu2021ReinforcementLF}, reinforcement learning is used to interact with an emotion recognizer to improve discriminability in the GST weight distribution. While the aforementioned approaches were successful in generating emotionally expressive speech of high quality, they were heavily reliant on emotion-labeled datasets. In this paper, we extend the exploration of GST in  E-TTS without the reliance on explicit emotion labels. 

\begin{figure*}[t]
    \centering
    \scalebox{0.65}
    {\includegraphics{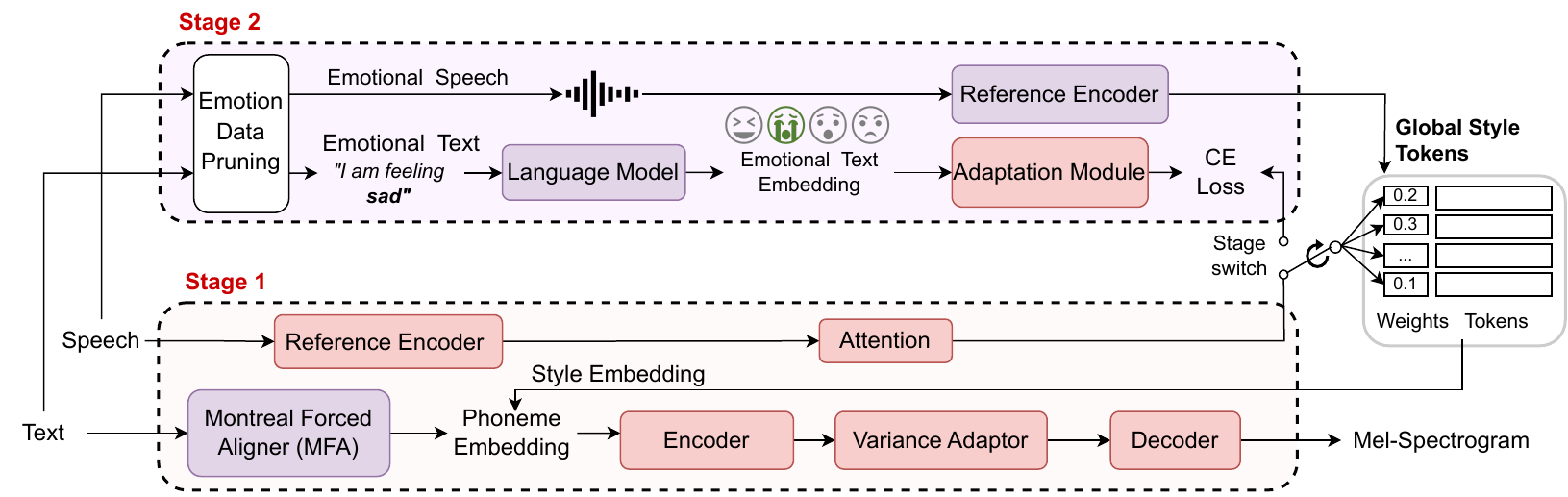}}
    \caption{Proposed TEMOTTS framework with stage I and stage II. Red modules contain trainable weights and purple modules contain fixed weights; CE loss represents Categorical Cross Entropy loss.}
    \label{framework}
\end{figure*}

Many E-TTS frameworks have been trained on acted datasets where the spoken content and intended emotion are mismatched \cite{Shin2022TextdrivenES}. Consequently, these models often struggle to capture conversational speaking styles effectively. Furthermore, several E-TTS models \cite{zhao2023emotion, Cai2020EmotionCS} require additional inputs such as reference audio and/or emotion labels during inference, resulting in speech that may not faithfully convey the emotional subtleties present in the input text. This limitation affects the scalability and significantly reduces the potential use cases of the TTS system. To address these challenges, our framework learns to capture and generate appropriate emotions solely based on the inherent emotional characteristics of the language, thus enhancing the system's robustness and adaptability.


In real life, our spoken words and affective prosody have a natural correlation, but this aspect is often overlooked when constructing TTS datasets and models. With the increasing capabilities of language models \cite{Devlin2019BERTPO,brown2020language} to understand deep semantic information from text, we are motivated to use this text awareness to connect the emotional context in the text to emotional prosody in speech. 

In this paper, we propose an E-TTS framework based on FastSpeech2, trained in two stages to directly learn emotion representations from emotion-unlabeled training data. We utilize style tokens to construct an emotional style space and a fine-tuned BERT model to explore this space effectively. 
The main contributions of the paper are summarized as follows:

\begin{itemize}

\item We present a novel two-stage training for an E-TTS model, eliminating the necessity for emotion labels during training;
\item We establish emotional text-awareness by establishing a direct connection between the linguistic and affective spaces using a language model; and
\item We present an inference strategy that enables the accurate synthesis of emotional speech, reflecting the emotion embedded within the text, all without the requirement for any auxiliary inputs. This is one of the first works to harness the emotional interconnection between our spoken content and their emotive delivery - for emotional speech synthesis.
\end{itemize}

\section{Related work}
\subsection{Towards using fewer emotion labels for E-TTS}
The scarcity of extensive emotion-labeled datasets poses significant limitations on E-TTS research. This limitation arises due to the high cost of labeling emotions for the extensive data required by TTS models \cite{kim2020glow}.
The demand for high-quality speech data in TTS restricts the direct utilization of datasets from speech emotion recognition (SER) research \cite{busso2008iemocap}. In light of these challenges, we are motivated to explore other ways to obtain emotional information to condition the TTS model. This problem of reducing the number of emotion annotations has been addressed through various approaches. In \cite{wu2019end}, a semi-supervised method was employed to enhance the interpretability of GST tokens in E-TTS. Similarly, in \cite{he2022improve}, an expressive dataset was leveraged to bolster the training of their graph neural network-based framework, integrating semi-supervised learning. Furthermore, \cite{Cai2020EmotionCS} introduced a cross-domain SER that predicted soft labels for the TTS dataset and utilized these labels to condition the GST style token weights.

In this paper, we omit the use of emotion labels and,  instead, emphasize a text-aware approach that leverages the inherent emotional connection within audio-text pairs.

\subsection{Text-aware TTS}
Beyond employing texts solely for learning alignments, exploring their utilization for acquiring prosodic representations constitutes an intriguing subfield in TTS research. Several expressive TTS frameworks, such as TP-GST \cite{Stanton2018PredictingES} and MsEmoTTS \cite{lei2023msstyletts} have explored text-awareness by learning representations for linguistic prosody. 
EMSpeech \cite{Cui2021EMOVIEAM}, on the other hand, directly aims to predict emotion labels from the text at inference by training on a dataset with human-labeled emotion annotations.

Incorporating language models like BERT \cite{kenton2019bert} into TTS has been another avenue of research. These models have been used to enhance prosody \cite{Chen2021SpeechBE, xiao2020improving} and learn emotional representations to condition the TTS model \cite{Mukherjee2022TextAE,wu22e_interspeech}. Recent models have also attempted to reduce the modality gap between style text inputs and reference speech by aligning the two modalities during training \cite{Kim2021ExpressiveTU, Shin2022TextdrivenES}. InstructTTS \cite{yang2023instructtts}, on the other hand, investigates text-awareness in two roles: as an input and as a style tag. With the increasing popularity of prompt encoders in TTS \cite{Guo2022PromptttsCT}, there is also a growing need for prompt-labeled datasets that need laborious manual annotations.

More recently, GPT-3 \cite{brown2020language} has been utilized to train a language model-based emotion prediction network \cite{Yoon2022LanguageME} for E-TTS. In contrast to previous studies that focused on training emotional speech synthesis using labeled data and incorporating auxiliary inputs during inference, our goal is to address the clear research gap between models that require these specific conditions for emotional speech synthesis and expressive TTS models that can operate without such constraints. We believe that language model based emotion prediction represents the future direction for advancing E-TTS systems. This aspect stands as a central focus of our paper.

\section{TEMOTTS}
We propose TEMOTTS, a text-aware E-TTS framework designed to directly acquire emotional speaking styles from textual input. Our choice of Fastspeech2 as the backbone is motivated by its explicit modeling of pitch, duration, and energy, resulting in the generation of expressive speech, as well as its efficiency in terms of both training and inference due to parallelism. It is worth noting that the underlying concept of TEMOTTS can be implemented within any encoder-decoder framework. 

TEMOTTS comprises two training stages: Stage I constructs a style space using GST and trains Fastspeech2 modules. In Stage II, an adaptation module maps emotional text embeddings from a language model to corresponding emotional style representations in the style space. We note that multi-step training strategies are shown to be effective in modeling alignment and prosody-related information separately \cite{liu2021expressive}. Since large, emotion-labelled data is hard to acquire, multi-step approaches allow for the independent modeling of the two tasks \cite{sivaprasad21_interspeech,kosgi2022empathic}. Our proposed two-stage training strategy is inspired by these methods, which have proven to be effective in addressing prosody-related challenges in E-TTS. 

\begin{figure*}[!h]
    \centering
    \scalebox{0.46}{\includegraphics{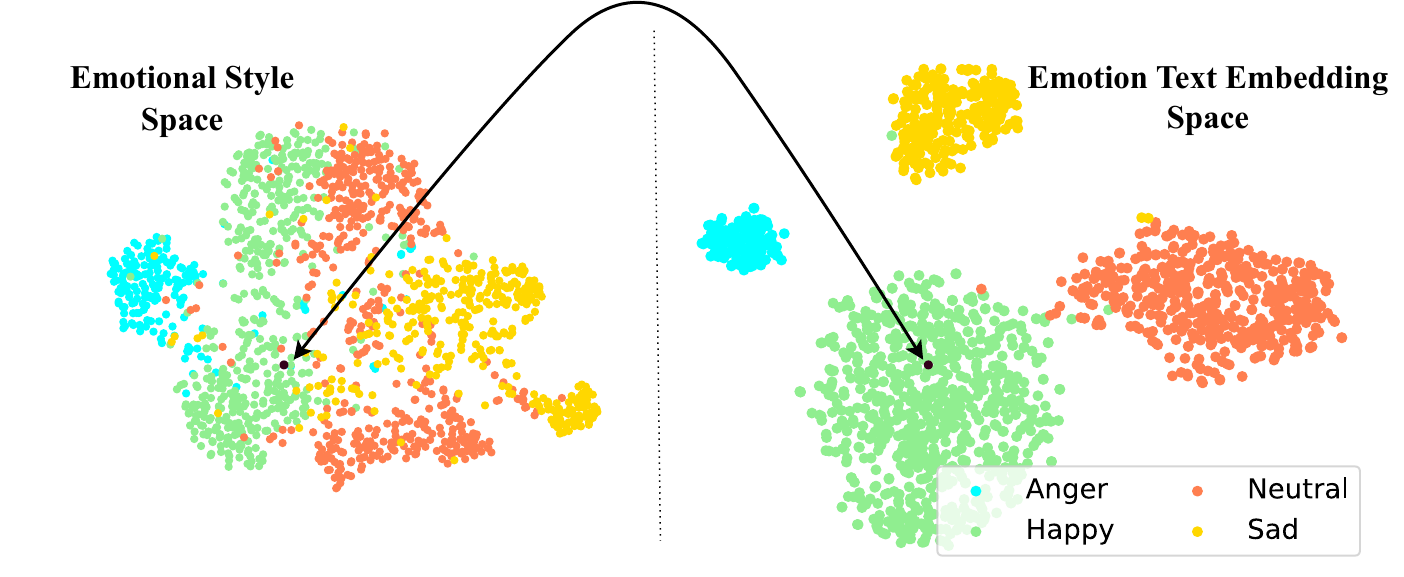}}
    \caption{Visualization of emotional style space and emotional text embeddings space with a conceptual link.}
    \label{Emotion_correlation}

\end{figure*}
\subsection{Stage I: Constructing an emotional style space}
In Stage I, we train the acoustic modules including a text encoder, variance adaptors, a decoder, and a GST network, as shown in Fig. \ref{framework}. The encoder converts phoneme embeddings into a phoneme-hidden sequence, while the variance adapter introduces pitch, duration, and energy variations. The decoder then translates this hidden sequence into the mel-spectrogram output.

The GST network, composed of a reference encoder and style embeddings, conditions the TTS model's encoder for various speaking styles. We specifically use GST because it constructs a style space in an unsupervised manner and provides a means to directly navigate this space using GST combination weights.

Since we use emotional training data, we hypothesize that the model implicitly learns speaking styles that are representative of emotions \cite{Um2019EmotionalSS}.  The style encoder is trained to decompose the input speech into a select set of basis vectors which can represent various speaking styles. This happens in an unsupervised manner and helps the base FastSpeech2 framework explore the latent space of emotions.
\subsection{Stage II: Exploration of style space with language model}
\label{sec:emotionpruning}

Stage II consists of three modules, as illustrated in Fig. \ref{framework}:
\\
\textbf{Emotion Data Pruning}: To capture emotional correlations between speech-text pairs, it is crucial to select utterances with substantial emotional context. To achieve this, we use a fine-tuned DistilRoBERTa-base language model \footnote{https://huggingface.co/michellejieli/emotion\_text\_classifier}, to predict emotion class probabilities from text. For each example, we calculate the dominant emotion using this model. If the probability of the dominant emotion class exceeds an experimentally chosen threshold $P_{th}$, we include the example in the training dataset for Stage II.
\\
\textbf{Emotion Text Embedding}: To represent emotional content in text, we extract emotion text embeddings using mean pooling over the last layer of the fine-tuned DistilRoBERTa-base language model.\\
\textbf{Adaptation Module}: We establish an emotional correlation between emotional context in text and emotional prosody in speech. In Fig.         \ref{Emotion_correlation}, we utilize t-SNE \cite{van2008visualizing} to visualize the emotional style space created by GST weight distribution in the emotion data-pruned dataset, alongside the space constituted by emotional text embeddings. A symbolic line connects two representative points from each space, illustrating the relationship between the emotional style space and the emotional text embedding space. Notably, Fig. \ref{Emotion_correlation} reveals a clustering similarity between the spaces. Motivated by this observation, we introduce an adaptation module capable of mapping emotional text embeddings from the language model to their corresponding emotional style representations. The emotion text embeddings are fed as input into this module and the GST weights, predicted by the trained reference encoder, serve as target values. This mapping is treated as a multi-class classification task, optimized using a categorical cross-entropy loss function. It is worth noting that we incorporate soft labels during training since the goal of stage II is to predict GST weights, which inherently range from 0 to 1.

\subsection{Inference}
\begin{figure}[t]
    \centering
    \scalebox{0.8}{\includegraphics{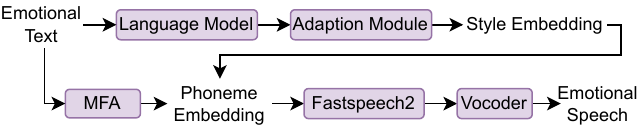}}
    \caption{Inference strategy of TEMOTTS.}
    \label{inference}
\end{figure}

During inference, TEMOTTS only requires text as input, as shown in Fig. \ref{inference}. The text is processed by the language model to generate an emotional text embedding. The adaptation layers predict GST weights, giving the model access to the emotion style space constructed by style tokens. The style embedding, derived from a weighted linear combination of these tokens, conditions the phoneme embedding module in stage I to synthesize text-aware emotional speech.

\begin{figure*}[!ht]
\centering
\subcaptionbox{VITS}{\includegraphics[width=0.25\textwidth]{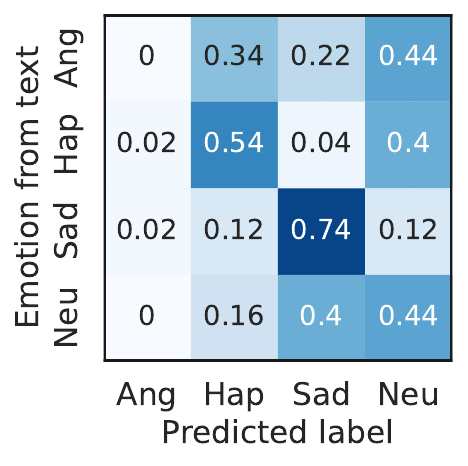}}%
\hspace{1cm}
\subcaptionbox{Fastspeech2}{\includegraphics[width=0.25\textwidth]{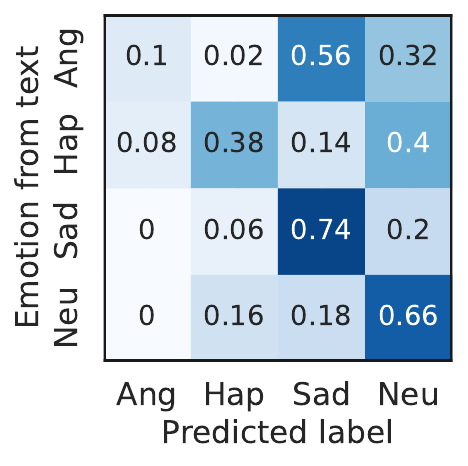}}%
\hspace{1cm}
\subcaptionbox{TEMOTTS}{\includegraphics[width=0.25\textwidth]{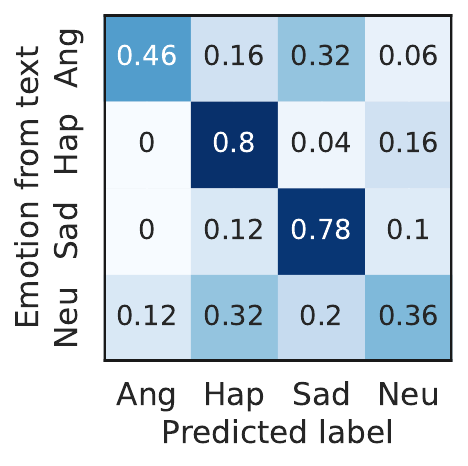}}%
\caption{SER confusion matrices of 4 emotion categories.}
\label{SER_fig}
\end{figure*}
\section{Experiments}
In this section, we conduct both objective and subjective evaluations to assess the performance of our proposed method, TEMOTTS, in terms of emotional accuracy and naturalness. 

We train TEMOTTS stage I using a combination of two datasets: LJSpeech \cite{ljspeech17} and an emotional TTS dataset \footnote{https://github.com/skit-ai/emotion-tts-dataset}\cite{Dalmia_Emotional_TTS_Dataset_2021}, both sampled at 22.05kHz. The emotional TTS dataset comprises high-quality recordings of nine emotions by an English female speaker. Unlike most acted emotional TTS datasets, this data is spoken in a natural style and has a high correlation between linguistic content and emotional speaking style. Although we train TEMOTTS with a recorded emotional TTS dataset, the model is designed to leverage any conversational speech dataset that has sufficient emotional variation. We believe that the acquisition of such data is more viable than manual annotation of emotional speech datasets. We partition our dataset into training, validation, and test sets with a ratio of 18:1:1.
For stage II, we use emotion data pruning described in Section \ref{sec:emotionpruning} to select a subset of the emotional dataset 
 \cite{Dalmia_Emotional_TTS_Dataset_2021}.
 For evaluation, we create two test sets using sentences generated by GPT-3 \cite{brown2020language}. The first set comprises of randomly generated sentences for assessing Word Error Rate (WER) and Character Error Rate (CER). The second set consists of sentences enriched with high-emotion content, designed to evaluate the emotional text-awareness of the models.

\subsection{Experimental setup}

We convert the raw waveform into a mel-spectrogram, configuring the hop size to 256 and making use of 80 mel bins. To extract phoneme duration, we utilize Montreal Forced Alignment \cite{mcauliffe2017montreal}. The adaptation module consists of eight fully connected dense layers of sizes - 772, 600, 500, 400, 300, 200, 100, 50, 40 and 16 with ReLU activation. We follow the     
ESPNet2\footnote{https://github.com/espnet/espnet} framework to implement other Fastspeech2-based modules and the reference encoder in GST network.   A pre-trained Parallel WaveGAN \cite{yamamoto2020parallel} \footnote{https://github.com/kan-bayashi/ParallelWaveGAN} trained on LJSpeech is used as the vocoder, while the remaining components are trained from scratch. 
We utilize the Adam optimizer with a Noam scheduler, using the following parameters: $\beta_1$ = 0.9, $\beta_2$ = 0.999, $\epsilon$ = 10e-9, 4000 warm-up steps and set the learning rate to 10e-4. The model with the best performance on the validation set is selected for evaluation after training. 

We set the emotion data pruning parameter $P_{th}$ to 0.7 after experimenting with values between 0.5 and 0.9.


\subsubsection{Choice of baselines:}
To ensure a fair comparison with the proposed TEMOTTS method, we consider baselines that are trained under the same constraints -
\begin{enumerate}
    \item without emotion labels; and
    \item inference without any auxiliary inputs.
\end{enumerate}
With the lack of such E-TTS frameworks, we consider Fastspeech2 \cite{ren2021fastspeech} and VITS \cite{kim2021conditional} for comparison. Both FastSpeech2 and VITS are expressive TTS frameworks that learn low-level prosody modeling from the training data. The training of Fastspeech2 is identical to stage I of our proposed method. For VITS, we train the model \footnote{VITS code: https://tts.readthedocs.io/en/latest/models/vits.html} with the same data as the proposed model. 
We also note that using style encoder-based TTS models would require additional inputs such as reference speech or prompts and hence does not constitute a fair baseline. 
\subsubsection{Choice of number of style tokens:} 
The selection of the number of style tokens ($N_{tokens}$) within the style encoder is a critical hyperparameter in our approach. Experimentation reveals that increasing $N_{tokens}$ significantly reduces the model's capacity to generate emotionally expressive speech while setting $N_{tokens}$ too low constrained the emotional style space. After thorough experimentation with $N_{tokens}$ values of 8, 16, 32, and 64, we find that $N_{tokens}$ = 16 provides the best results.

\subsection{Objective evaluation}
We calculate CER and WER to assess the intelligibility of the synthesized speech. Table \ref{table:CER} presents CER and WER results obtained from the Wav2Vec 2.0 Large ASR model\footnote{https://huggingface.co/facebook/wav2vec2-large-960h-lv60-self}. These results indicate that our proposed method, with its text-awareness, enhances the intelligibility of synthesized speech compared to the baselines.

To assess the capability of the model to synthesize emotional speech, we train an SER by finetuning Wave2Vec2.0 \cite{baevski2020wav2vec} on the emotional TTS dataset.  We synthesize 50 sentences each for the four emotion classes - anger, happiness, neutral, and sadness - using GPT3 \cite{brown2020language}. We evaluate the ability of the models to synthesize the expected emotion from text using the trained SER. We treat the dominant emotion present in the text as the true label. Confusion matrices from the SER (Fig. \ref{SER_fig}) clearly illustrate that the proposed model demonstrates a higher level of confidence in synthesizing emotional speech that conveys the intended emotion from text, in contrast to the baseline models.

\begin{table}[t]
\caption{Objective and subjective evaluation results, where MOS results have 95\% confidence intervals.}\label{table:CER}
\centering
\scalebox{0.8}{\begin{tabular}{c c c cc c}
\toprule
\multirow{2}{*}{Method} & \multirow{2}{*}{CER(\%) } & \multirow{2}{*}{WER(\%)} & \multicolumn{2}{c}{BWS}       & \multirow{2}{*}{MOS} \\ \cline{4-5}
                  &                          &                          & Best(\%)        & Worst(\%)    &                      \\ 
                  \hline
Ground Truth      & 1.68                     & 7.32                     & \multicolumn{2}{c}{-}         & -                    \\ 
VITS\cite{Wang2017TacotronTE}              & 8.22                     & 18.57                    & 4.20            & 79.52         &     3.02  $\pm$ 0.07                \\
Fastspeech2\cite{ren2021fastspeech}       & 6.88                     & 18.34                    & 14.76          & 18.57         &     3.68 $\pm$ 0.06                 \\
\textbf{TEMOTTS}  & \textbf{4.08}            & \textbf{13.11}           & \textbf{80.95} & \textbf{1.90} & \textbf{3.75  $\pm$ 0.06}\\ \hline
\end{tabular}}
\end{table}

\begin{figure}[!h]
\centering
\subcaptionbox{Happy text: \textit{"I am excited for the new football season to start!"}}{\includegraphics[width=0.3\textwidth]{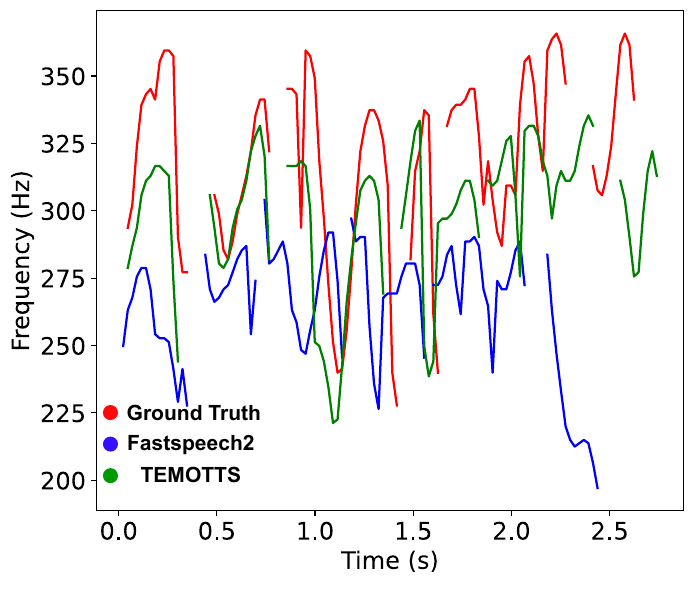}}%
\\
\subcaptionbox{ Sad text: \textit{"I am so sad these days."}}{\includegraphics[width=0.3\textwidth]{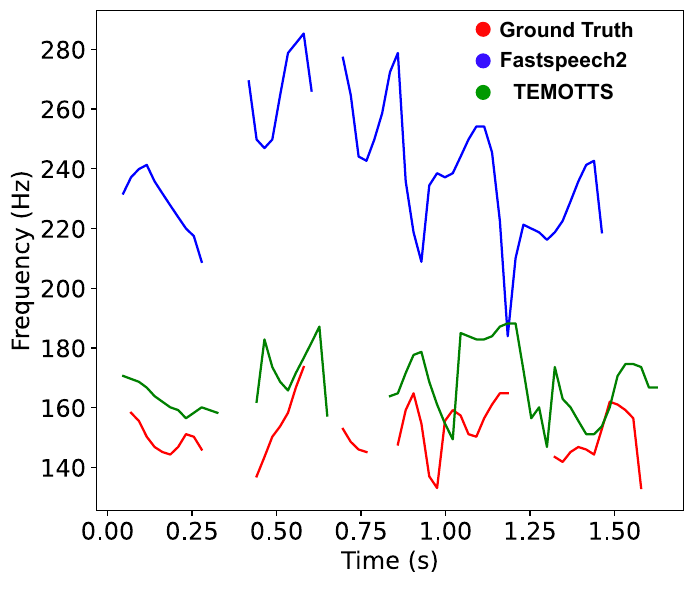}}%
\caption{Pitch contours of speech synthesized by TEMOTTS and Fastspeech2 along with ground truth.}
\label{contours}
\end{figure}


\subsubsection{Comparing pitch contours:}
In Fig. \ref{contours}, we plot the pitch contour of two different emotional sentences. We clearly observe that a) TEMOTTS is capable of modeling pitch more accurately with respect to the ground truth, and b) TEMOTTS has a higher dynamic range in pitch. This shows the effectiveness of our model in terms of expressiveness and ability to convey emotions accurately.
\subsection{Subjective evaluation}
We conduct listening experiments with 15 subjects to assess speech quality and the ability to synthesize text-aware emotional speech, using a total of 90 utterances for the tests.

In the Mean Opinion Score (MOS) evaluation \cite{Wang2017TacotronTE}, participants rate speech quality using a five-point scale for 15 speech samples generated by each model, with text generation performed randomly by GPT-3\cite{brown2020language}.
As indicated in Table. \ref{table:CER}, the proposed TEMOTTS generates speech with a higher level of naturalness than both baselines, achieving a MOS score of 3.75. We believe that the ability of TEMOTTS to be emotionally aware while synthesizing speech leads to better naturalness.

To assess the emotional text-awareness of the model, we conduct a Best-Worst Scaling (BWS) test. We used GPT-3 to generate 15 sentences with high emotion polarity. We then synthesize these sentences using our method and baselines, and listeners are tasked with selecting the best and worst samples based on how accurately the speech conveys the emotion in the text. Table. \ref{table:CER} shows that TEMOTTS outperforms both baselines in terms of synthesizing text-emotion accurate speech, receiving the highest best votes \textit{(80.95\%)} and the lowest worst votes \textit{(1.9\%)}.

In both subjective and objective evaluations, the proposed method shows the clear advantage of being emotionally text-aware. Even though the baselines Fastspeech2 and VITS are trained to learn appropriate low-level prosody features for the input text, we clearly show that this does not necessarily constitute the learning of emotional prosody. Hence, the proposed method outperforms these baselines, especially in the cases where the spoken content is emotional in nature.


\section{Conclusion}
In this study, we introduce TEMOTTS, an E-TTS framework that overcomes the challenges posed by human-annotated emotion labels and the intricate nature of emotional prosody learning. TEMOTTS relies on text awareness to acquire emotional styles, eliminating the need for explicit emotion labels during training and auxiliary inputs during inference. Our experiments showcase significant improvements in emotional accuracy, naturalness, and intelligibility compared to baselines. This work presents a novel approach to enhancing the emotional richness of speech synthesis by connecting spoken content and its expressive delivery.
\bibliographystyle{IEEEbib}
\bibliography{ref}

\end{document}